\documentclass[10pt,conference]{IEEEtran}

\IEEEoverridecommandlockouts

\usepackage{cite}
\usepackage{amsmath,amssymb,amsfonts}
\usepackage{algorithm, algorithmic}
\usepackage{graphicx}
\usepackage{textcomp}
\usepackage{xcolor}
\usepackage{comment}
\usepackage{booktabs}
\usepackage{multirow}
\usepackage{tabularx}
\usepackage{mathtools}
\usepackage{caption}
\usepackage[normalem]{ulem}
\usepackage{soul}

\def\BibTeX{{\rm B\kern-.05em{\sc i\kern-.025em b}\kern-.08em
    T\kern-.1667em\lower.7ex\hbox{E}\kern-.125emX}}
\begin{document}

\title{Edge Learning via Federated Split Decision Transformers for Metaverse Resource Allocation}

\author{\IEEEauthorblockN{Fatih Temiz, Shavbo Salehi, and Melike Erol-Kantarci, \textit{Fellow, IEEE}}
\IEEEauthorblockA{\textit{School of Electrical Engineering and Computer Science, University of Ottawa, Ottawa, Canada} \\
Emails: \{ftemi033, ssale038, melike.erolkantarci\}@uottawa.ca}}

\maketitle

\begin{abstract}
Mobile edge computing (MEC) based wireless metaverse services offer an untethered, immersive experience to users, where the superior quality of experience (QoE) needs to be achieved under stringent latency constraints and visual quality demands. To achieve this, MEC-based intelligent resource allocation for virtual reality users needs to be supported by coordination across MEC servers to harness distributed data. Federated learning (FL) is a promising solution, and can be combined with reinforcement learning (RL) to develop generalized policies across MEC-servers. However, conventional FL incurs transmitting the full model parameters across the MEC-servers and the cloud, and suffer performance degradation due to naive global aggregation, especially in heterogeneous multi-radio access technology environments. To address these challenges, this paper proposes Federated Split Decision Transformer (FSDT), an offline RL framework where the transformer model is partitioned between MEC servers and the cloud. Agent-specific components (e.g., MEC-based embedding and prediction layers) enable local adaptability, while shared global layers in the cloud facilitate cooperative training across MEC servers. Experimental results demonstrate that FSDT enhances QoE for up to 10\% in heterogeneous environments compared to baselines, while offloading nearly 98\% of the transformer model parameters to the cloud, thereby reducing the computational burden on MEC servers.
\end{abstract}


\section{INTRODUCTION}
The rapid evolution of the metaverse has ushered in a new era of immersive experiences that extend far beyond entertainment, reaching into education, healthcare, culture, and global social interactions. It offers immersive services via wireless virtual reality (VR), where the service is tailored to improve the quality of experience (QoE) of its users and overcomes barriers stemming from physical constraints of wired experiences. However, latency becomes a significant issue when high-resolution field-of-view (FoV) rendering tasks are performed solely on head-mounted displays (HMDs) with limited hardware capabilities\cite{9411714}. Therefore, mobile edge computing (MEC) has garnered significant attention for offloading computationally intensive tasks to the edge network \cite{9411714} or at least collaborating with HMDs \cite{11003161}, aiming for lower latency and higher visual qualities to avoid motion sickness. Furthermore, human vision is known to be inherently hierarchical, comprising central, paracentral, and peripheral regions that perceive visual information with varying levels of detail and sensitivity \cite{8824804}. Gaze-aware streaming systems can exploit this hierarchical property to allocate MEC resources more effectively, using head and eye tracking to deliver high-fidelity rendering to the regions of the FoV where users focus most. 

Reinforcement learning (RL) has demonstrated strong potential for complex, dynamic and time-sensitive environments, providing adaptive resource allocation 
and scheduling in VR streaming \cite{9411714}. Nevertheless, online RL algorithms such as deep deterministic policy gradient (DDPG) can struggle to converge to the optimal policy when exploration is insufficient \cite{11080254}. Digital twin (DTW) technologies have been proposed to enhance online RL frameworks by enabling safer emulation and training in the digital space while mitigating desynchronization through real-time updates to the RL policy with continual learning \cite{10486201}. Although DTW can support RL, its development and 
maintenance are costly due to the need for extensive data collection and storage of historical data in a buffer. Constructing a high-fidelity DTW that replicates both static and dynamic contexts in real time would inevitably incur significant overhead. To address the shortcomings of online RL, offline RL approaches, such as decision transformers (DTs), have emerged as promising alternatives by using historical trajectories to achieve more stable and sample-efficient learning \cite{NEURIPS2021_7f489f64}, and it is an emerging paradigm for wireless network optimization, especially for intent-driven network management \cite{11112781} and resource allocation problems \cite{10839243}. 
Beyond the online–offline distinction, federated learning (FL) can be combined with reinforcement learning (federated reinforcement learning, FRL) by enabling distributed training of generalized policies on distributed nodes or MEC servers, while addressing data scarcity on individual MEC servers without requiring raw data sharing \cite{10464431}. 

Recent studies have explored the application of FL and FRL within MEC-enabled metaverse scenarios, incorporating FoV rendering–based QoE models \cite{Mudi2025ICC,11080254}. 
In \cite{Mudi2025ICC}, authors design a network slicing framework in an open radio access network environment and train the FRL model on multiple agents to optimize power control and physical resource block allocation. Their FoV rendering-based QoE-driven approach demonstrates significant improvements in delay reduction and user experience compared to benchmark algorithms. Also, the authors in \cite{11080254}, present FedPromptDT, a federated offline RL algorithm based on prompt-guided DT and a custom QoE model, and show that it outperforms both online algorithms and centralized DT training using the eye tracking dataset. 
 
Besides incorporating FL into the mobile edge using distributed MEC resources, split learning 
has a crucial role in the 6G era of edge artifical intelligence (AI). By hierarchically partitioning the model, split learning enables more efficient use of dispersed computing resources across the network edge, ultimately improving overall resource utilization \cite{10529950}. In \cite{10778352}, the authors propose a personalized FRL framework for heterogeneous edge content caching networks. Their approach combines a multi-head deep Q-network with layer-wise personalized federated training, where the model is split to balance global knowledge sharing with local personalization. This design mitigates the limitations of naive global aggregation in FL, which often leads to homogenized caching strategies and degraded performance across diverse MEC servers. However, the above system still do not benefit from offline RL paradigms that rely on advanced transformer architectures.

Recently, federated split training of transformer-based architectures was explored in \cite{NEURIPS2021_ceb05951}, 
particularly for vision transformers and later for DTs \cite{10651270}. These architectures are used  mainly for task-agnostic training and show strong performance in heterogeneous environments, allowing different agents to specialize while still benefiting from shared global representations. For example, in vision applications, one agent may handle classification through its local embedding and prediction layers, while another focuses on segmentation, yet both continue to draw benefits from the shared global transformer decoder. This setup enables making better use of distributed data and, at the same time, gives each agent the flexibility to adapt to its own task or environment without sacrificing the advantages of cooperative learning.

Building on this line of work, this paper proposes a Federated Split Decision Transformer (FSDT)-based resource allocation framework for metaverse applications, addressing challenges in MEC data scarcity, online RL instability, naive global aggregation, and inefficient full-model exchanges in FL.
The FSDT approach combines FL and split learning
for training the DT algorithm, where the model is split between the MEC-server and the cloud during training. This design maintains local adaptability through agent-type-specific embedding and prediction layers and benefits from global coordination via the shared transformer in the cloud. The key contributions of this paper are as follows:

\begin{enumerate}
    \item We propose a cooperative edge AI framework via model partitioning, where the computationally intensive global transformer decoder is deployed in the cloud, while distributed MEC servers host the embedding and prediction layers, enabling domain-specific customization. 
    \item We formulate the problem of FSDT-based resource allocation for metaverse in heterogeneous multi-RAT networks, and make use of a public VR eye-tracking dataset.
    \item We conduct extensive experiments to evaluate multiple trade-offs, including online versus offline RL algorithms and FRL under both training modes, thereby providing a comprehensive benchmark of the proposed FSDT framework across diverse settings. 
\end{enumerate}

\begin{figure*}[htbp]
    \centering
    \includegraphics[width=1\textwidth]{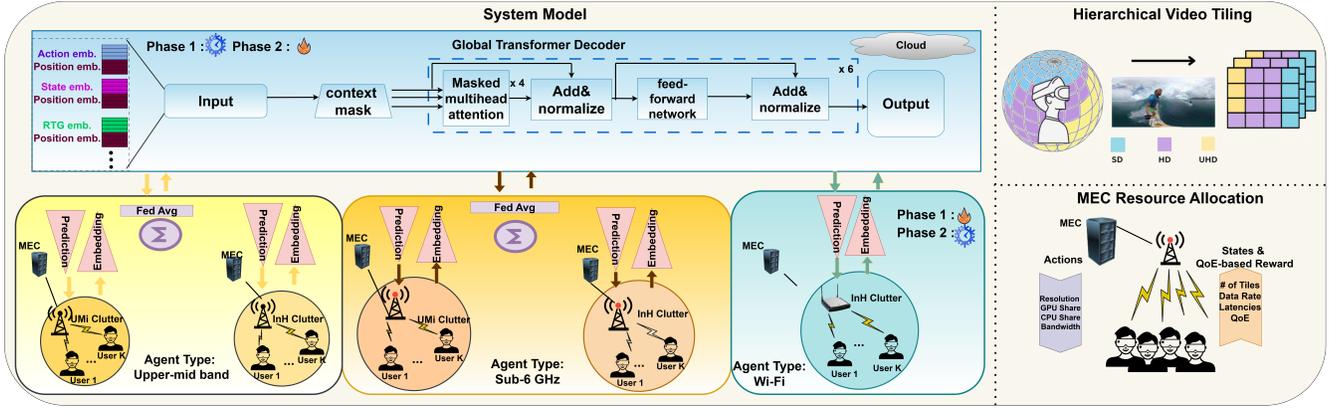}
    \caption{System Model}
    \label{fig:SystemModel}
\end{figure*}

\section{SYSTEM MODEL AND PROBLEM FORMULATION}
\subsection{System Model}
In this work, we consider a heterogeneous multi-RAT MEC system that renders and serves immersive 360$^{\circ}$ video streams according to requests from wireless VR users with HMDs as can be seen in Fig. \ref{fig:SystemModel}. The system consists of collocated base stations (BSs), each connected to an MEC server, that allocate resources including GPU rendering cycles, CPU encoding cycles, bandwidth (BW), and resolution ratios to ensure the timely delivery of high-quality video to VR users.
BSs are grouped into $K$ different agent types according to the employed access technologies: \textit{Upper-mid band (UMB)}, \textit{Sub-6 GHz}, and \textit{Wi-Fi}. Within each agent type (i.e., domain), BSs operate under different clutter environments such as Urban-Micro (UMi) and Indoor-Hotspot (InH), reflecting diverse propagation conditions. 
Additionally, to support adaptive streaming, we follow the approach in \cite{10486201} where each $360^\circ$ video is projected onto a 2D plane and divided into 
tiles (e.g., $4 \times 4$) as illustrated in Fig. \ref{fig:SystemModel}. For user $k$ at time slot $t$, only the tiles in FoV \(\mathcal{F}_k(t)\) are delivered at high quality, while peripheral tiles use reduced resolution. Three quality levels $q \in \{1,2,3\}$ corresponding to standard definition (SD), high definition (HD), and ultra high definition (UHD) are used for hierarchical vision.
The tiles are encoded per group of picture (GoP) of $F$ frames, and in each time slot, one 
GoP is transmitted. 
Then, for each user \(k\), the data size at the quality level \(q\) is
\begin{equation}
d_{k,q}(t) = M_{k,q}(t)\,\beta_{k,q}(t)\,F,
\end{equation}
where \(M_{k,q}(t)\) denotes the number of tiles in quality \(q\) for user \(k\) (with \(\sum_q M_{k,q}(t) = M_k(t)\); e.g., \(3+9+4=16\) in Fig.~\ref{fig:SystemModel}), \(\beta_{k,q}(t)\) is the per-tile bitrate at quality level \(q\), and the total data size is \(D_k(t) = \sum_q d_{k,q}(t)\) (bits). In addition, we model \(\beta_{k,q}(t) = B_{q,\max}\, r_{k,q}(t)\), where \(r_{k,q}(t)\) is the resolution ratio assigned to user \(k\) at the quality level \(q\). This GoP tiling design prioritizes FoV resolution while reducing overall transmission and computation costs.

\subsubsection{Communication Model}
The large-scale channel gain is modeled using the close-in (CI) path loss model with log-normal shadowing \cite{7434656}. The path loss at distance $d$ (in meters) and carrier frequency $f_c$ (in Hz) is given as:
\begin{equation}
PL(d,f) \; [\text{dB}] = 20 \log_{10}\!\left(\frac{4 \pi f d_0}{c}\right) 
+ 10 \alpha \log_{10}\!\left(\frac{d}{d_0}\right) + X_{\sigma},
\end{equation}
where $d_0=1$ m is the reference distance, $\alpha$ is the path loss exponent (PLE), $X_{\sigma} \sim \mathcal{N}(0,\sigma^2)$ models log-normal shadowing, and $c$ is the speed of light. 

The achievable downlink (DL) data rate of user $k$ associated with BS $i$ is modeled as:
\begin{equation}
\label{eq:rate}
R_k = B_k \log_2 \left( 1 + \frac{P_k g_k^2 h_k}{N_0 B_k \cdot \mathrm{NF}} \right),
\end{equation}
where $B_k$ is the allocated BW, $P_k$ is the transmit power, $g_k$ is the small-scale Rayleigh fading coefficient, $h_k$ PL in linear scale
, $N_0$ is the thermal noise power spectral density, and $\mathrm{NF}$ is the receiver noise figure (in linear scale). Accordingly, the transmission latency is given by:
\begin{equation}
    T^{\text{comm}}_k(t) = \frac{D_k(t)}{C_r R_k},
\end{equation}
where $R_k$ is the achievable DL data rate, $C_r$ is the compression factor,
and $D_{k}(t)$ denotes the total data size of task from user $k$ at time slot $t$.

\subsubsection{Computation Model}
At the MEC server, the processing of each GoP for user $k$ at time slot $t$ consists of GPU-based rendering and CPU-based encoding computational tasks \cite{11003161}.

\begin{itemize}
    \item \textbf{GPU-based Rendering:}  
    Let $\gamma^{\text{gpu}}$ denote the GPU cost in pixel/tile at quality level $q$, and $f_k^{\text{gpu}}$ the GPU capacity (pixels/s) allocated to the user $k$.  Each pixel consists 3 channels (i.e., RGB) totaling 24 bits per pixel. 
    Then the total number of pixels is
    \begin{equation}
        P_k(t) = F\sum_q M_{k,q}(t)\,p_{k,q},
    \end{equation}
    where $p_{k,q}$ is the number of pixels per tile at quality $q$, the rendering latency is
    \begin{equation}
        T^{\text{rend}}_k(t) = \frac{\gamma^{\text{gpu}} P_k(t)}{f_k^{\text{gpu}}}.
    \end{equation}
    \item \textbf{CPU-based Encoding:}  
    Let $\gamma^{\text{cpu}}$ denote the CPU cost in cycles/bit, and $f_k^{\text{cpu}}$ the CPU capacity (cycles/s).  
    Given the total encoded data size $D_k(t)$ in bits,
    the encoding latency is
    \begin{equation}
        T^{\text{enc}}_k(t) = \frac{\gamma^{\text{cpu}} D_k(t)}{f_k^{\text{cpu}}}.
    \end{equation}
\end{itemize}

Therefore, the end-to-end latency for user $k$ is
\begin{equation}
    T^{\text{tot}}_k(t) = T^{\text{rend}}_k(t) + T^{\text{enc}}_k(t) + T^{\text{comm}}_k(t).
\end{equation}

\subsubsection{Quality of Experience}
In VR streaming, QoE is affected by several factors such as tiling, stall, latency, and resolution switching.  
We incorporate three major components to measure QoE precisely, namely latency, tiling based video quality, and resolution switch. Given these considerations, the QoE for user $k$ in time slot $t$, could be formulated as follows:

\begin{subequations}\label{eq:QoE}
\begin{align}
\text{QoE}_k &= \left(1 - \tfrac{T^{\text{tot}}_k}{T_{\text{th}}}\right) S_k 
                - \lambda R_{\text{amp},k} ,
                \tag{\theparentequation}\label{eq:QoE-main} \\
S_k &= \sum_q \tfrac{w_q M_q}{M_{\text{total}}} 
        \cdot \log\!\left(1 + \tfrac{\beta_q}{\beta_{\text{th},q}}\right),
        \label{eq:QoE-S} \\
R_{\text{amp},k} &= \sum_q \big| r_q^{(t)} - r_q^{(t-1)} \big|,
        \label{eq:QoE-R}
\end{align}
\end{subequations}
where latency in eq. (\ref{eq:QoE-main}) reflects network conditions, video quality component in eq. (\ref{eq:QoE-S}) captures perceptual aspects beyond traditional QoS metrics \cite{10486201} through the weight factor $w_q \in \{\text{low}:1,\; \text{medium}:2,\; \text{high}:3\}$, and $R_{\text{amp},k}$ in eq. (\ref{eq:QoE-R}) is the resolution switching amplitude which is a crucial metric in wireless VR for user experience as validated in \cite{8647729} with mean opinion score correlation.

\subsection{Problem Formulation and RL Framework}

\subsubsection{Problem Statement}
In this subsection, the formulation for QoE maximization across all users explained. The MEC resource allocation problem can be formulated as follows:

\begin{subequations}\label{eq:optimization}
\begin{equation}\label{eq:optimization_obj}
\max_{\{f_k^s, b_k^s, g_k^s, r_{k,q}\}} \;
\sum_{k} \mathrm{QoE}_{k},
\tag{\theparentequation}
\end{equation}
\begin{equation}\label{eq:optimization_alloc}
\text{s.t.}\quad
\sum_k f_k^{sh} \le 1,\quad
\sum_k b_k^{sh} \le 1,\quad
\sum_k g_k^{sh} \le 1,\quad
\forall t
\end{equation}
\begin{equation}\label{eq:optimization_rates}
0.125 < r_{\mathrm{low}} < 0.25 < r_{\mathrm{med}} < 0.5,\quad
0.75 < r_{\mathrm{high}} < 1,
\end{equation}
\begin{equation}\label{eq:optimization_qoe}
\mathrm{QoE}_{k,t} \ge \mathrm{QoE}_{k,\min},\quad \forall k,t
\end{equation}
\begin{equation}\label{eq:optimization_bounds}
0 \le f_k^{sh},\, b_k^{sh},\, g_k^{sh},\, r_{k,q} \le 1,
\end{equation}
\end{subequations}
where in eq. (\ref{eq:optimization_alloc}), $f_k^{\text{sh}}$, $b_k^{\text{sh}}$, and $g_k^{\text{sh}}$ denote the CPU, BW, and GPU fractions respectively, and 
eq. (\ref{eq:optimization_bounds}) ensures valid CPU, GPU, and BW allocations by restricting all allocation and rate variables to the unit interval. Also, eq. (\ref{eq:optimization_rates}) establishes quality limits \cite{10486201}, and eq. (\ref{eq:optimization_qoe}) penalizes per-user QoE violations.

\subsubsection{MDP Formulation}
The RL agent's Markov Decision Process (MDP) is defined as follows:
\paragraph{State Space} At time $t$ over $K$ users, the system state captures per-user DL data rates, delay components, tile distributions, and QoE:
\begin{equation}
    \mathbf{s}_t = \left[ R_k, T_k^{\text{c}}, T_k^{\text{e}}, T_k^{\text{r}}, T_k^{\text{tot}}, \{M_{k,q}\}_{q \in \{\text{high}, \text{med}, \text{low}\}}, \text{QoE}_k \right]_{k=1}^{K},
\end{equation}
where $R_k$ is the DL transmission rate
according to eq. (\ref{eq:rate}), $T_k^{(\cdot)}$ are the DL communication, encoding, rendering, and total delays, respectively, and $M_{k,q}$ denotes the number of tiles assigned to the quality level $q$.

\paragraph{Action Space} At time $t$, the system chooses an action that jointly determines the allocation of computational ($f_k^{sh},g_k^{sh}$) and communication ($b_k^{sh}$) resources, as well as the resolution ratios ($r_{k,q}$) for all $K$ users:
\begin{equation}
    \label{eq:action_space}
    \mathbf{a}_t = \left[ f_k^{\text{sh}}, b_k^{\text{sh}}, g_k^{\text{sh}}, \{r_{k,q}\}_{q \in \{\text{high}, \text{med}, \text{low}\}} \right]_{k=1}^{K},
\end{equation}
For instance, with 4 users, the action space and state space dimensions are 24 and 36, respectively.
\paragraph{Reward Function}
The reward function balances user QoE with fairness and constraint violations:
\begin{equation}
r_t = \sum_{k=1}^{K} \text{QoE}_{k,t} - \omega_{\text{fair}} \mathcal{F}_t - \omega_{\text{viol}} \mathcal{V}_t,
\end{equation} 
where $\mathcal{F}_t = \sigma(\{f_{k,t}\}_{k=1}^K) + \sigma(\{b_{k,t}\}_{k=1}^K) + \sigma(\{g_{k,t}\}_{k=1}^K)$ represents the fairness penalty based on resource allocation variance across users, $\mathcal{V}_t$ denotes QoE threshold violations at time $t$, and $w_{fair}$ and $w_{viol}$ represents the system degree of sensitivity towards fairness and violations.

\section{FSDT-based METAVERSE RESOURCE ALLOCATION}

\subsection{FSDT}
The architecture of FSDT is built upon the DT paradigm
, reformulated to operate in a federated environment \cite{10651270}.
DT formulates RL as a sequence modeling problem where decision-making is learned through autoregressive modeling rather than an explicit value-function or policy optimization task and can be expressed as follows. For user $k$, contextual MDP is defined as $(\mathcal{S}_k, \mathcal{A}_k, \mathcal{P}_k, \mathcal{R}_k)$ with states $s_t^k \in \mathcal{S}_k$, actions $a_t^k \in \mathcal{A}_k$, rewards $r_t^k = \mathcal{R}_k(s_t^k, a_t^k)$, and $\mathcal{P}(s'|s,a)$ is the transition dynamics. 
The model uses undiscounted RTG, i.e., the sum of future rewards, $\hat{R}_t^k = \sum_{t'=t}^{T} r_{t'}^k$ to enable autoregressive trajectory modeling. 
This yields the following trajectory representation, which can be effectively used for autoregressive training and sequence generation:

\begin{equation}
\tau^k = \bigcup_{t=1}^T \left( \hat{R}_t^k,\, s_t^k,\, a_t^k \right),
\end{equation}
where $T$ denotes the trajectory length.
Please note that the goal of offline RL is to learn a policy that maximizes the expected return 
$\mathbb{E}\!\left[\sum_{t=1}^T r_t\right]$ in the MDP. 
Then, embedding layer maps these RTGs, states, and actions into 
hidden space, enriched with temporal embeddings (i.e., positional embedding) $\pi(t)$:
\begin{equation}
u_{x,t}^{k_n} := \pi(t) \oplus \varphi_x(x_t^{k_n}), \quad x \in \{r,s,a\}.
\end{equation}
These embeddings passed through stacked \emph{masked causal self-attention blocks} as illustrated in Fig. \ref{fig:SystemModel}, in which the decoder architecture closely follows the GPT architecture \cite{NEURIPS2021_7f489f64}. 
The decoder on the cloud produces contextualized token outputs:
\begin{equation}
(v_{r,t}^{k_n},\, v_{s,t}^{k_n},\, v_{a,t}^{k_n}) 
= G\!\left( \{u_{r,\tau}^{k_n},\, u_{s,\tau}^{k_n},\, u_{a,\tau}^{k_n}\}_{\tau=t-L}^t \right).
\end{equation} 
where $L$ is the context length (i.e., number of RTG-state-action tuples to predict the next action). The prediction module $P_{k_n}$ maps the contextualized representation to the next action using sigmoid activation. Then, the output $\hat{a}_t^{k_n}$ is scaled to the valid action range. Building upon this DT formulation, FSDT extends the approach to a federated environment.

\subsection{Training Procedure}
\label{subsec:training_procedure}
FL and split learning paradigms are among the most common distributed learning paradigms in communication systems. In these approaches, each client (e.g., MEC-server) develops a local model using their own datasets. Clients send only model updates (e.g., gradients or embeddings) to a central aggregator (e.g., cloud), which merges them into a global model and shares back with the clients for further refinement.

\begin{algorithm}[h]
\caption{VR-FSDT Training }
\label{alg:FSDT}
\textbf{Input:} 
Domain-global models $\{ g_0^k \}_{k=1}^K$, 
server model $v_0$, 
client-local models $w_0^{k,n}$,
communication rounds $R$. \\
\textbf{Output:} Final Domain-global $\{G_R^k\}_{k=1}^K$, server model $v_R$
\begin{algorithmic}[1]
\FOR{$r = 1$ to $R$}
    \STATE /* Phase 1: update embedding-prediction only */
    \FOR{$k = 1$ to $K$}
        \FOR{$n = 1$ to $N_k$}
            \STATE Initialize $w_r^{k,n} \gets g_{r-1}^k$
            \STATE $w_r^{k,n} \gets \text{UpdateLocal}(w_r^{k,n}, v_{r-1})$
        \ENDFOR
        \STATE $\bar{w}_r^k \gets \text{FedAvg}(w_r^{k,1},\ldots,w_r^{k,N_k})$
        \STATE $G_r^k \gets \text{DomainUpdate}(\bar{w}_r^k, v_{r-1})$
    \ENDFOR
    \STATE $v_r \gets v_{r-1}$
    \STATE /* Phase 2: update server transformer only */
    \FOR{$k = 1$ to $K$}
        \FOR{$n = 1$ to $N_k$}
            \STATE $v_r \gets \text{TrainServer}(G_r^k, v_r)$
        \ENDFOR
    \ENDFOR
\ENDFOR
\end{algorithmic}
\end{algorithm}
The training of FSDT proceeds in two alternating stages across $R$ communication rounds (see Alg.~\ref{alg:FSDT}) and follows the principles of FL and split learning. At each round, clients first adapt their local modules to personalize to their own data, while keeping the server transformer decoder fixed (lines 2-10). Then, the server trains the shared transformer decoder 
using all local data (lines 11-18). The objective is to minimize the mean squared error (MSE) between predicted actions and ground-truth actions from the DDPG trajectory dataset. This design balances local adaptability (Phase 1) and global coordination (Phase 2).

\textbf{Phase 1:}
At round $t$, the server sends the latest domain-global parameters to all clients of type $k$. Each client $n \in \{1,\ldots,N_k\}$ updates its local embedding and prediction modules $(e_t^{k,n}, p_t^{k,n})$ while keeping the server transformer $G_{t}$ fixed:
\begin{equation}
\label{eq:lossfunction}
\min_{\{e_t^{k,n},\,p_t^{k,n}\}} 
\sum_{k=1}^K \sum_{n=1}^{N_k} 
\text{MSE} \!\left( y_t^{k,n},\, 
p_t^{k,n}\big(G_{t}(e_t^{k,n}(x_t^{k,n}))\big)\right).
\end{equation}
Then, client updates are aggregated, i.e., federated averaging (FedAvg), within each agent type (e.g., across clutters of UMB domain):

\begin{equation}
e_t^k = \frac{1}{N_k}\sum_{n=1}^{N_k} e_t^{k,n}, 
\quad
p_t^k = \frac{1}{N_k}\sum_{n=1}^{N_k} p_t^{k,n}.
\end{equation}
\textbf{Phase 2:}
With $(e_t^k, p_t^k)$ frozen, the server updates its transformer $G_t$ using embeddings from all client types and minimizes the same objective as in Eq. \ref{eq:lossfunction} but with respect to $G_t$ only.
A single cloud-shared decoder is trained jointly across MECs, while embedding and prediction layers remain domain-specific. The decoder is replicated post-training and combined with the domain-specific layers, resulting in UMB, Sub-6 GHz, and WiFi models.

\section{SIMULATION SCENARIOS AND RESULTS}

\subsection{Experimental Settings and Baselines:}
In our simulations, we have considered five heterogenous MEC environments with multi-RAT setting, each serving four users, whose positions are randomly distributed within a square region, where the side length corresponds to the maximum distance specified in Table \ref{tab:EnvParams}. The table also reports the key clutter parameters, including BW, carrier frequency $f_c$, PLE ($\alpha$), shadowing variance $\sigma$, and $P$ 
under non-line of sight 
conditions \cite{bazzi2025upper}.
Moreover, the dataset is obtained from \cite{8578657} that contains 200+ YouTube videos (4K, 25fps, 20-60s). The dataset is temporally split into 80\% and 20\% for training and test, respectively. Per-frame 2D eye coordinates determine tile quality levels using Chebyshev distance (Fig.~\ref{fig:SystemModel}).
All the results in the following subsections presented as the average over 3 seeds 
\begin{table}[!htbp]
\captionsetup{labelfont=normal,labelsep=colon} 
\centering
\scriptsize
\caption{multi-RAT Parameters}
\label{tab:EnvParams}
\begin{tabular}{|l|c|c|c|c|c|c|}
\hline
\textbf{RAT/Clutter} & \textbf{BW} & \textbf{$f_c$} & \textbf{PLE} & \textbf{$\sigma$} & \textbf{Dist.} & \textbf{$P$} \\
 & \textbf{[MHz]} & \textbf{[GHz]} & \textbf{$\alpha$} & \textbf{[dB]} & \textbf{[m]} & \textbf{[dBm]} \\
\hline
\hline
UMB/UMi & 200 & 6.75 & 2.56 & 6.53 & 230 & 33 \\
UMB/InH & 200 & 6.75 & 2.72 & 9.21 & 65 & 33 \\
\hline
Sub6GHz/UMi & 100 & 2.90 & 2.90 & 2.90 & 230 & 30 \\
Sub6GHz/InH & 100 & 2.90 & 3.10 & 6.50 & 65 & 30 \\
\hline
Wi-Fi/InH & 160 & 2.40 & 2.52 & 5.75 & 65 & 24 \\
\hline
\end{tabular}
\vspace{-5pt}
\end{table}
\subsubsection{Baselines} 
We evaluated five algorithms, including four baselines and FSDT as shown in Table \ref{tab:baselines}, where, the prefixes C- and F- denote centralized and federated, respectively. 
Offline approaches (i.e., C-DT, F-DT and FSDT) 
leverage pre-collected trajectories generated using C-DDPG. FSDT is trained for 20 rounds, where each round consists of 100 client gradient timesteps (Phase 1) followed by 100 server gradient timesteps (Phase 2).
During testing, all of the methods are tested over 100 episodes over the same test dataset.

\begin{table}[h]
\centering
\caption{Comparison of baselines and the proposed FSDT.}
\label{tab:baselines}
\setlength{\tabcolsep}{3pt} 
\renewcommand{\arraystretch}{0.9} 
\begin{tabular}{lcccc}
\toprule
\textbf{Algorithm} & \textbf{Train. Mode}  & \textbf{Aggregation} & \textbf{Exchanged Info.} \\
\midrule
C-DDPG\cite{10486201}  & Centralized       & N/A         & N/A \\
F-DDPG                 & Federated         & All layers  & Critic Params \\
C-DT\cite{NEURIPS2021_7f489f64} & Centralized  & N/A & N/A \\
F-DT                   & Federated       & All layers  & Full Params \\
FSDT                   & Fed.-Split      & Local layers & Embeddings \\
\bottomrule
\end{tabular}
\end{table}

\vspace{-10pt}

\begin{table}[h]
\centering
\caption{FSDT and MEC parameters.}
\label{tab:fsdt_mec_params}
\setlength{\tabcolsep}{3pt}
\renewcommand{\arraystretch}{1.0}
\begin{tabular}{@{}l c p{3cm} c@{}}
\toprule
\textbf{FSDT parameter} & \textbf{Value} & \textbf{MEC parameter} & \textbf{Value} \\
\midrule
\texttt{n\_heads}        & \texttt{4}      & $\gamma_{q}^{\text{cpu}}$ ($q$=1,2,3) & 200, 190, 180 \\
\texttt{n\_blocks}       & \texttt{6}      & $F$, $M$                             & 16, 16 \\
\texttt{context\_length} & \texttt{50}     & $T_{th}$ (ms), $QoE_{\min}$          & 200, 4 \\ 
\texttt{optimizer}       & \texttt{AdamW}  & $C_r$, $NF$ (dB)                     & 300, 7 \\
\texttt{learning\_rate}  & $10^{-4}$       & $f^{\text{CPU}}$ (cyc/s),$f^{\text{GPU}}$ (px/s)              & $15\times10^{9}$ \\
\texttt{weight\_decay}   & $10^{-4}$       &       $w_{fair}$, $w_{viol}$       & 2, 2 \\
\texttt{hidden\_dim}     & \texttt{256}    &     $\beta_{max,q}$ (Mbits)             & 12.4, 1.4, 0.5 \\
\bottomrule
\end{tabular}
\end{table}

\vspace{-2pt}
\subsection{Results}
\begin{figure}[t]
    \centering
    \includegraphics[width=0.88\columnwidth]{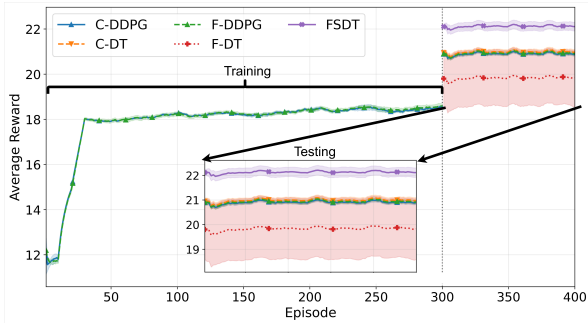}
    \caption{Average reward versus episode for baselines and FSDT
    }
    \label{fig:reward}
    \vspace{-15pt}
\end{figure}

In terms of the reward, Fig.~\ref{fig:reward} presents the moving average reward of FSDT compared with the baselines, showing training results for the online RL algorithms 
and test results for all algorithms.
The results demonstrate that FSDT consistently outperforms its baselines with up to 10\% higher rewards by integrating offline RL, FL, and split learning. These paradigms complement one another to provide key advantages, particularly improved learning 
through strategic sampling, mitigating the scarcity of local MEC data, and overcoming the limitations of naive global aggregation by model partition, respectively. 
 F-DDPG and C-DDPG perform quite similarly during both training and testing. FSDT and C-DT show less variance across seeds compared to F-DT and show nearly 10\% and 5\% greater performance.

\begin{figure}[t]
    \centering
    \begin{minipage}[t]{0.5\columnwidth}
        \centering
        \includegraphics[width=\linewidth]{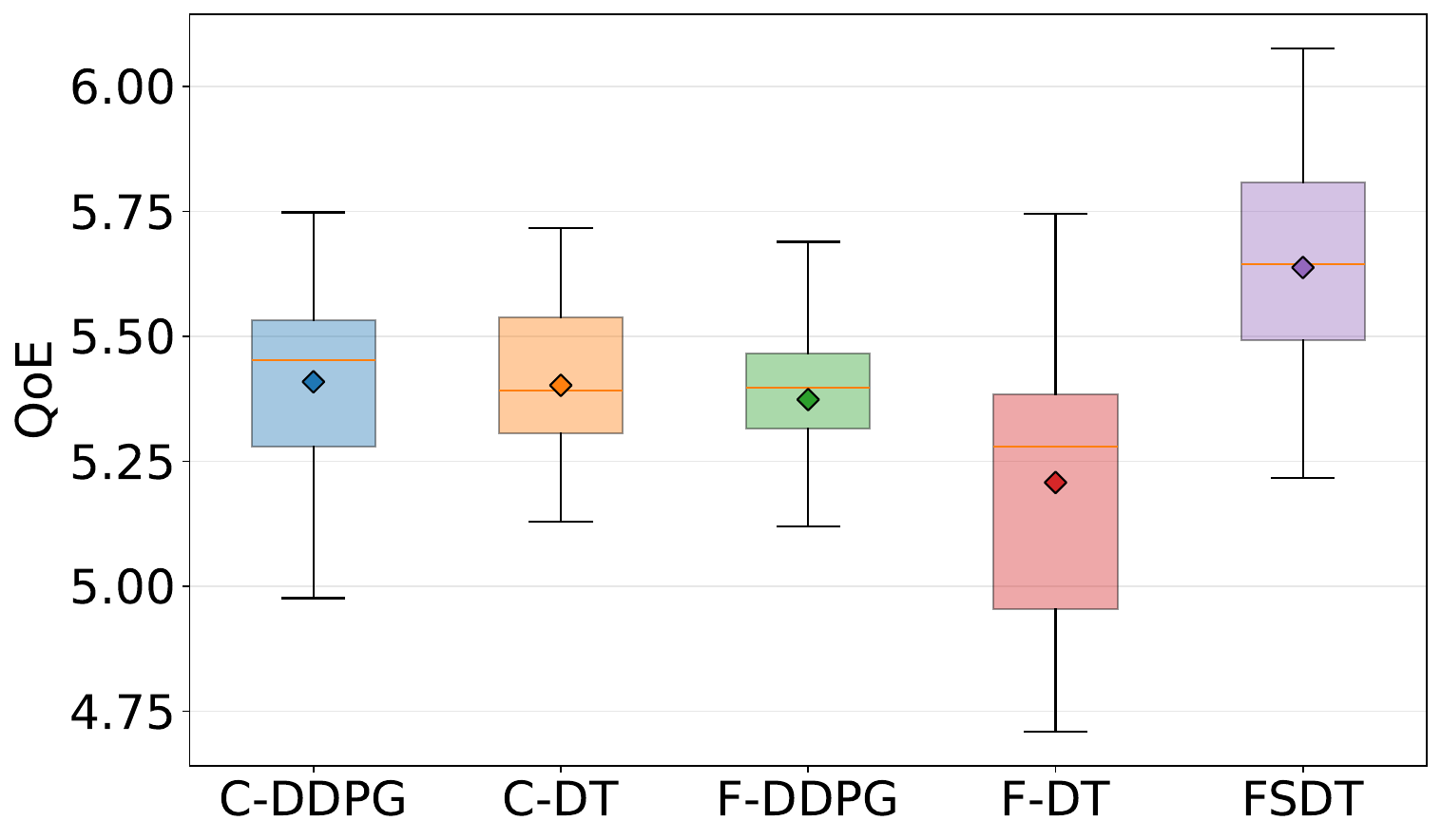}
        \caption{QoE Comparison}
        \label{fig:QoE}
        \vspace{-15pt}
    \end{minipage}%
    \hfill
    \begin{minipage}[t]{0.5\columnwidth}
        \centering
        \includegraphics[width=\linewidth]{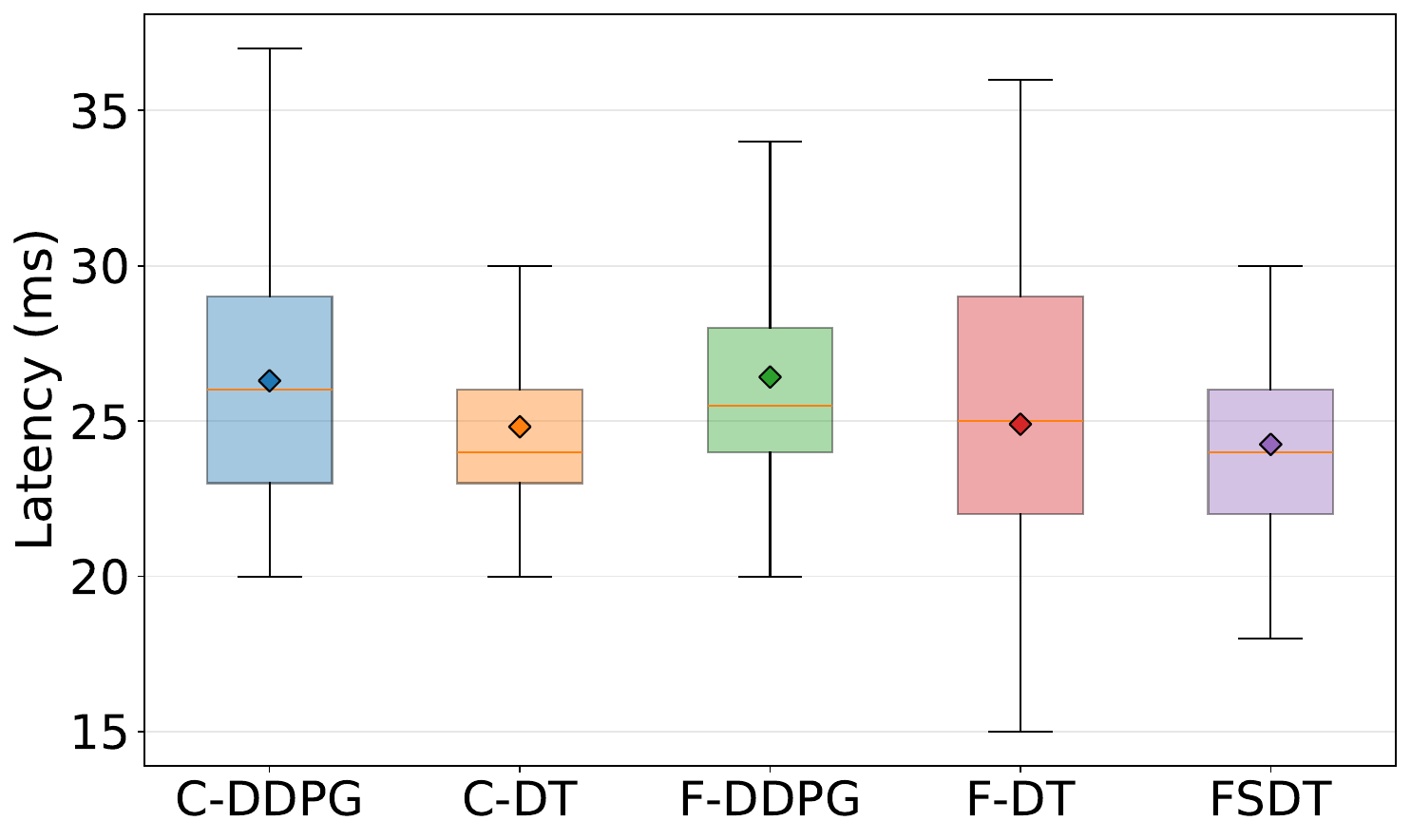}
        \caption{Latency Comparison}
        \label{fig:latency}
        \vspace{-15pt}
    \end{minipage}
\end{figure}

Figs. \ref{fig:QoE} and \ref{fig:latency} show the distributions of user QoE and latency across all MECs and users. The boxplots indicate the median, interquartile range, overall spread, and mean (diamond). In terms of QoE, Fig. \ref{fig:QoE} depicts that most schemes achieve a median QoE between 5.2 and 5.5. The proposed FSDT method attains the highest performance, with a median QoE of about 5.6 and an upper whisker reaching 6.07, clearly outperforming all baselines. By contrast, F-DT has the lowest median QoE (e.g., 5.1) and exhibits wider variability, indicating less stable performance. Moreover, the comparison between C-DDPG and F-DDPG suggests that conventional FL does not enhance online RL in heterogeneous settings, as updating shallow critic networks provides limited benefit for collaborative learning. Also, the same trend can be seen in comparison between C-DT and F-DT where traditional FL (e.g., FedAvg) does not improve QoE. 
In contrast, FSDT improves C-DT 
since each  MEC learn from the intra-domain MECs by FedAvg of embedding and prediction layers and also from the inter-domain MECs by shared transformer decoder.

In terms of latency, Fig. \ref{fig:latency} shows latency where all schemes lie in the range of 20–30 ms on average. C-DT and FSDT achieve the lowest median latencies, around 22–23 ms, with FSDT showing a tighter spread, indicating more consistent performance. C-DDPG and F-DDPG show higher medians around 26 ms and wider whiskers. 

\vspace{-5pt}
\subsection{Complexity Analysis}
Table~\ref{tab:model_layers} summarizes FSDT model subnetworks. The transformer decoder dominates with 98.79\% of parameters, keeps computation in the cloud and does not need to be frequently exchanged across MECs. 
This design reduces MEC server load, training time, and enhances privacy by sharing only latent representations of actions, states, and RTGs. The training communication cost of conventional FL and FSDT can be expressed as
\begin{equation}
\scalebox{0.9}{$
T_{FL}=2RP_{tot}, \quad
T_{FSDT}=2BR(F+G)+2R(P_e+P_p),
$}
\end{equation}
where $P_{tot}$, $P_e$, and $P_p$ denote the number of total, embedding, and prediction parameters, respectively, $B$ is the batch size (e.g., 64), and $F$ and $G$ represent the transmitted features and gradients at the split point \cite{NEURIPS2021_ceb05951}. Depending on model hyperparameters, this architecture can reduce overall communication across MECs.

\begin{table}[h!] 
\centering 
\caption{FSDT Split Details}
\label{tab:model_layers} 
\begin{tabular}{|l|c|c|c|} \hline \textbf{Metric} & \textbf{Decoder (Cloud)} & \textbf{Embed (Edge)} & \textbf{Pred (Edge)} \\ \hline \# of Params & 4,738,560 & 42,496 & 15,677 \\ \hline Size (MB) & 18.076 & 0.162 & 0.060 \\ \hline Memory (\%) & 98.79 & 0.89 & 0.33 \\ \hline 
\end{tabular}
\vspace{-5pt}
\end{table}
\vspace{-4pt}
\section{CONCLUSION}
In conclusion, the resource allocation framework based on FSDT for edge-cloud collaborative learning in heterogeneous multi-RAT environments demonstrates superior performance by combining the offline RL, FL and split learning. Compared to baseline, where the model is not split, 
it shows more consistent results benefiting both local layers and shared global transformer decoder. FSDT not only achieves higher QoE with lower latencies up to 10\%, but also improves the training efficiency by keeping computational burden on clouds and enhances privacy. 
Future work may explore edge-device collaboration or task-agnostic model development.
\vspace{-5pt}

\section*{ACKNOWLEDGEMENT}
This work has been supported by the NSERC Canada Research Chairs Program. We gratefully acknowledge late Dr. Han Zhang, a recent alumna of our research lab,  for helping us access the dataset used in this study. We dedicate our paper to her loving memory.

\vspace{-10pt}
\renewcommand\refname{References}
\addcontentsline{toc}{section}{References}
\bibliographystyle{IEEEtran}
\bibliography{ref}

\end{document}